\documentclass{emulateapj}

\def\gs{\mathrel{\raise1.16pt\hbox{$>$}\kern-7.0pt %
\lower3.06pt\hbox{{$\scriptstyle \sim$}}}}         %
\def\ls{\mathrel{\raise1.16pt\hbox{$<$}\kern-7.0pt %
\lower3.06pt\hbox{{$\scriptstyle \sim$}}}}         %

\slugcomment{}
\shorttitle{The AGN Contribution to the Mid-IR Emission of LIRGs}
\shortauthors{}

\begin{document}

\title{The AGN Contribution to the Mid-IR Emission of Luminous Infrared Galaxies}

\author{K. Brand\altaffilmark{1}, A. Dey\altaffilmark{1}, D. Weedman\altaffilmark{2}, V. Desai\altaffilmark{3}, E. Le Floc'h\altaffilmark{4}$^,$\altaffilmark{10}, B.~T. Jannuzi\altaffilmark{1}, B.~T. Soifer\altaffilmark{3},\altaffilmark{5}, M. J. I. Brown\altaffilmark{6},  P. Eisenhardt\altaffilmark{7}, V. Gorjian\altaffilmark{7}, C. Papovich\altaffilmark{4}$^,$\altaffilmark{8}, H.~A. Smith\altaffilmark{9}, S.~P. Willner\altaffilmark{9}, R.~J. Cool\altaffilmark{4}} 

\altaffiltext{1}{National Optical Astronomy Observatory, 950 North Cherry Avenue, Tucson, AZ 85726; brand@noao.edu} 
\altaffiltext{2}{Department of Astronomy, Cornell University, 610 Space Sciences Building, Ithaca, NY 14853}
\altaffiltext{3}{Division of Physics, Mathematics and Astronomy, California Institute of Technology, 320-47, Pasadena, CA 91125}
\altaffiltext{4}{Steward Observatory, University of Arizona, 933 North Cherry Avenue, Tucson, AZ 85721}
\altaffiltext{5}{Spitzer Science Center, California Institute of Technology, 323-6, Pasadena, CA 91125}
\altaffiltext{6}{Princeton University Observatory, Peyton Hall, Princeton, NJ 08544}
\altaffiltext{7}{Jet Propulsion Laboratory, California Institute of Technology, 169-327, Pasadena, CA 91109}
\altaffiltext{8}{Spitzer Fellow}
\altaffiltext{9}{Harvard Smithsonian Center for Astrophysics, 60 Garden Street, Cambridge, MA 02138}
\altaffiltext{10}{Associated to Observatoire de Paris, GEPI, 92195 Meudon, France}

\begin{abstract}

We determine the contribution of AGN to the mid-IR emission of luminous infrared galaxies (LIRGs) at $z>0.6$ by measuring the mid-IR dust continuum slope of 20,039 mid-IR sources. The 24 ${\rm \mu m}$ sources are selected from a {\it Spitzer/MIPS} survey of the NOAO Deep Wide-Field Survey Bo\"otes field and have corresponding 8 ${\rm \mu m}$ data from the {\it IRAC} Shallow Survey. There is a clear bimodal distribution in the 24 ${\rm \mu m}$ to 8 ${\rm \mu m}$ flux ratio. The X-ray detected sources fall within the peak corresponding to a flat spectrum in $\nu$f$_\nu$, implying that it is populated by AGN-dominated LIRGs, whereas the peak corresponding to a higher 24 ${\rm \mu m}$ to 8 ${\rm \mu m}$ flux ratio is likely due to LIRGs whose infrared emission is powered by starbursts. The 24 ${\rm \mu m}$ emission is increasingly dominated by AGN at higher 24 ${\rm \mu m}$ flux densities (f$_{24}$): the AGN fraction of the $z>0.6$ sources increases from ~9\% at f$_{24}\approx$0.35 mJy to 74$\pm$20\% at f$_{24}\approx$ 3 mJy in good agreement with model predictions. Deep 24 $\rm \mu m$, small area surveys, like GOODS, will be strongly dominated by starburst galaxies. AGN are responsible for $\sim$3-7\% of the total 24 ${\rm \mu m}$ background. 
\end{abstract}


\section{Introduction}

Luminous and Ultraluminous infrared galaxies (LIRGs) have huge infrared luminosities (L$_{\rm IR}\ge 10^{11}~{\rm L_{\odot}}$), corresponding to an extremely active phase of dust enshrouded star formation and/or AGN activity. They are under-luminous at rest-frame UV wavelengths because they are reprocessing and radiating much of their energy in the infrared (e.g., \citealt{san96}). LIRGs become an increasingly significant population at high redshifts (e.g., \citealt{lef04}) and are likely to dominate the luminosity density at $z > 1$, representing an important phase in the build-up of massive galaxy bulges and in the growth of their central super-massive black holes (SMBHs). To understand these processes, it is essential to separate the contribution of AGN and starbursts to the IR luminosity of LIRGs.

Recent work has shown how the {\it IRAC} color-color diagram and {\it MIPS} 24 ${\rm \mu m}$ to 8 ${\rm \mu m}$ color can be used to identify AGN-dominated LIRGs (\citealt{lac04}; \citealt{saj05}; \citealt{ste05}; \citealt{yan04}). In this paper we demonstrate how the 24 ${\rm \mu m}$ to 8 ${\rm \mu m}$ flux ratio ($\zeta\equiv{\rm log_{10}}[\nu f_\nu(24{\rm \mu m})/\nu f_\nu(8{\rm \mu m})]$) can be used to disentangle the contribution of AGN and starbursts to the total reprocessed mid-IR ($\approx 5-25~\rm \mu m$) emission as a function of 24 ${\rm \mu m}$ flux. The utility of the mid-IR slope as an AGN/starburst diagnostic was first demonstrated on samples of nearby {\it IRAS} galaxies (\citealt{san88}; \citealt{soi91}) and further developed by \citet{lau00}. The powerful UV and X-ray emission from an AGN can heat the surrounding dust to a range of temperatures up the sublimation temperature of $\approx$1500 K, resulting in a shallow mid-IR power-law continuum (e.g., \citealt{elv94}; \citealt{kla01}; \citealt{lut98}). In contrast, hot stellar sources in star-forming galaxies do not typically heat dust to these temperatures, resulting in a steeper mid-IR continuum slope. If the observed 24 ${\rm \mu m}$ and 8 ${\rm \mu m}$ flux densities are not heavily influenced by absorption and emission features, AGN-dominated LIRGs should have smaller values of $\zeta$ than starburst-dominated LIRGs. Cooler dust temperatures which are characterized by steeper mid-IR continua may also be expected in sources containing only a weak or heavily obscured AGN. We will return to these issues in Section~\ref{sec:dis}. In this work, we use the mid-IR slope as an AGN/starburst diagnostic applied to a large statistical sample of sources observed with the {\it Spitzer Space Telescope}.

\begin{figure*}
\plottwo{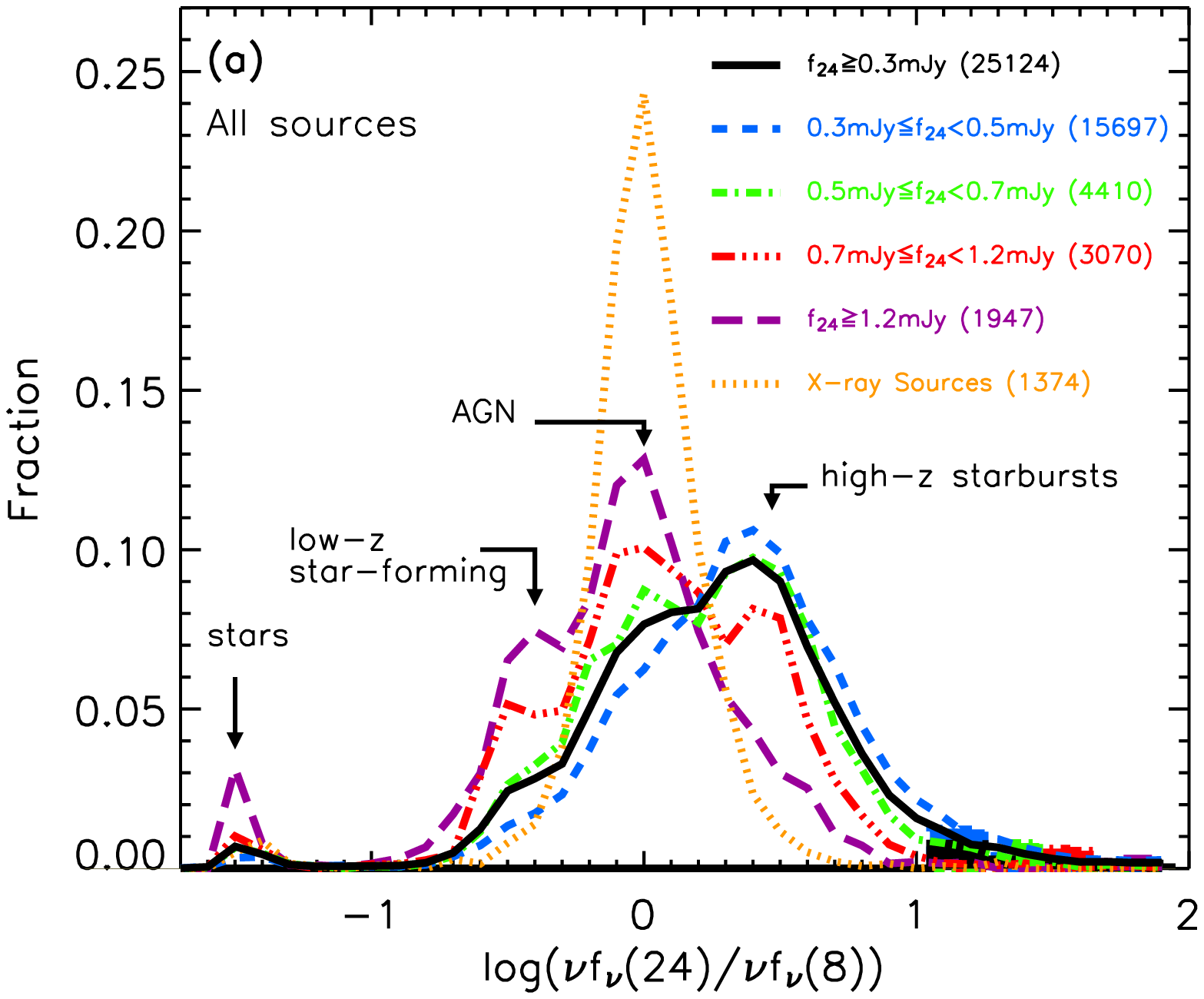}{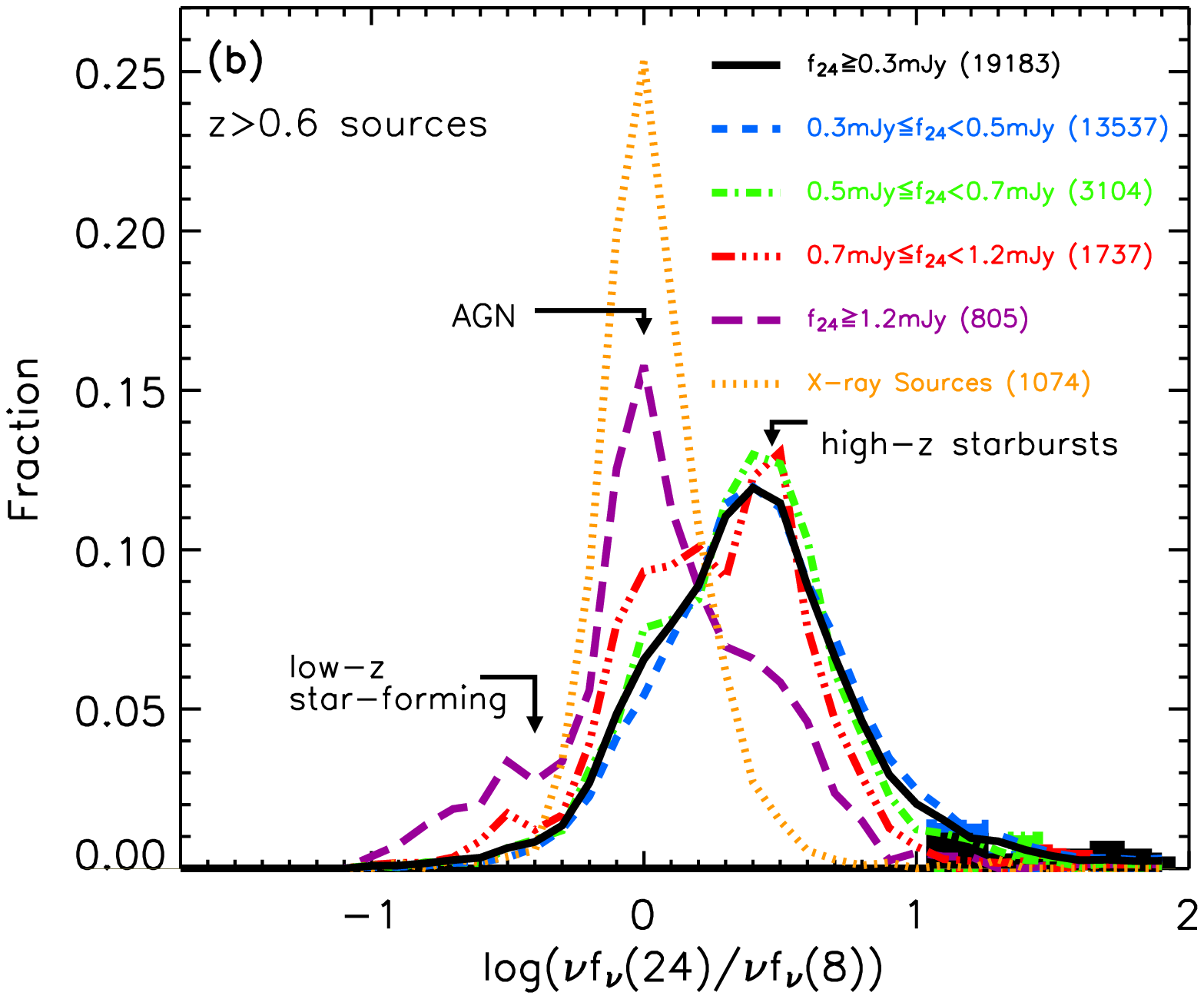}
\figcaption{\label{fig:f24_8} Histograms showing (a) $\zeta$ for all 24 ${\rm \mu m}$ sources, and (b) all 24 ${\rm \mu m}$ sources excluding those with $z\le 0.6$, split into f$_{24}$ bins. The histograms are for f$_{24}\ge 0.3$ mJy (all sources; solid black line), $0.3\le {\rm f}_{24}<0.5$ mJy (short-dashed blue line), $0.5\le {\rm f}_{24}<0.7$ mJy (dot-dashed green line), $0.7\le {\rm f}_{24}<1.2$ mJy (dot-dot-dot-dashed red line), and f$_{24}\ge 1.2$ mJy (long-dashed purple line). Also plotted is the $\zeta$ distribution of X-ray sources detected in the XBo\"otes survey (dotted orange line). The number of sources contributing to each histogram is listed in the figure. The shaded histograms show the $\zeta$ distribution of the 24 ${\rm \mu m}$ sources with no 8 ${\rm \mu m}$ counterparts assuming an 8 ${\rm \mu m}$ flux density limit of 3 times the locally calculated sky {\small RMS}.}
\end{figure*}

\section{Data}

The sample for this study was drawn from {\it Spitzer} observations of the NOAO Deep Wide-Field Survey Bo\"otes field (NDWFS; \citealt{jan99}). The 9.93 deg$^2$ region has been mapped by {\it Spitzer/MIPS} at 24 ${\rm \mu m}$ (f$_{24}$; comprising 28,391 sources down to a 5$\sigma$ depth of $\approx$0.3 mJy; version made on July 8th 2005). An 8.5 deg$^2$ region of the field has been mapped in all four {\it Spitzer/IRAC} bands by the {\it IRAC} Shallow Survey (\citealt{eis04}), reaching 5$\sigma$ depths at 8 ${\rm \mu m}$ of $\approx$0.076 mJy (for the fluxes measured in 6$^{\prime\prime}$ apertures). The 24 ${\rm \mu m}$ catalog was created by running a PSF fitting technique using {\small DAOPHOT}. In this paper, only the 25,681 24 ${\rm \mu m}$ sources whose positions overlapped with data in both the NDWFS and {\it IRAC} Shallow Survey are considered ($\approx$ 8.2 deg$^2$). Optical and infrared identifications were determined by matching the 24 ${\rm \mu m}$ catalog to the multi-wavelength catalogs using a 2$^{\prime\prime}$ matching radius. We used the 8 ${\rm \mu m}$ flux densities (f$_8$) measured in 6$^{\prime\prime}$ apertures corrected to the total flux density assuming a point source profile. In cases with no 8 ${\rm \mu m}$ catalog detection (1588 sources), f$_8$ was measured directly from the images; when no signal was found (557 sources), the f$_8$ limit (defined as 3 times the locally measured sky {\small RMS}), was used. In practice, the inclusion or exclusion of these sources does not significantly affect any of our conclusions. 

The Bo\"otes region has also been surveyed by the 5-ks {\it Chandra X-ray Observatory} ( XBo\"otes; \citealt{mur05}; \citealt{ken05}; \citealt{bra06}).
The AGN and Galaxy Evolution Survey (AGES; Kochanek et al.~in prep.; \citealt{bro05}) has also targeted the Bo\"otes field, providing spectroscopic redshifts for 98\% of the f$_{24}\ge0.3$ mJy 24 ${\rm \mu m}$ sources with $I\le21.5$ (optical point sources) and f$_{24}\ge0.5$ mJy sources with $I\le20.0$ (optical extended sources). As described in Section~3, we use this survey to exclude $z\le$0.6 sources in which strong PAH emission may fall into the observed 8 ${\rm \mu m}$ bandpass.

\section{Analysis}
\label{sec:analy}

We calculated $\zeta$ for each 24 ${\rm \mu m}$ source as a crude measure of the mid-IR continuum slope. This provides an estimate of the relative luminosities of the hot ($\approx$ 750 K at $z\approx$1) and warm ($\approx$ 250 K at $z\approx$1) dust components. Histograms of $\zeta$ for different f$_{24}$ bins are plotted in Figure~\ref{fig:f24_8}a. The distribution has 4 peaks, located at $\zeta\approx-1.5$, $\zeta\approx-0.4$, $\zeta\approx0$, and $\zeta\approx0.5$. As we discuss below, we identify these peaks with stars, low-redshift normal star-forming galaxies, AGN, and higher redshift starburst galaxies respectively. In addition, sources have progressively lower values of $\zeta$ at higher f$_{24}$. Although the sources with no 8 ${\rm \mu m}$ detection become a larger fraction at smaller f$_{24}$, they should only affect the tail of the distributions at $\zeta>1.0$ and will not significantly affect the location or width of either peak.

\subsection{Low redshift contaminants}

Strong PAH emission and silicate absorption can bias the broad-band flux density measurements of LIRGs. In low redshift ($z<0.3$) galaxies, strong 7.7 ${\rm \mu m}$ PAH emission can bias $\zeta$ to smaller values, rendering normal star-forming galaxies and AGN difficult to separate (e.g., \citealt{yan04}). To avoid this bias, we use existing spectroscopic redshifts from AGES and conservatively exclude the 5642 sources with $z\le0.6$ from our sample. Although AGES is $\approx$98\% complete for optically bright sources, optically faint ($I\ge 20.0$) low-redshift ($z\le0.6$) galaxies may remain in our sample. Figure~\ref{fig:f24_8}b shows that after exclusion of the AGES $z\le0.6$ sources, the $\zeta\approx-1.5$ and $\zeta\approx-0.4$ peaks become only a small residual and the distribution of the remaining 20,039 sources becomes largely bimodal. By fitting the $\zeta$ peaks with Gaussian profiles, we estimate the residual contamination by low-redshift galaxies in our $z>0.6$ sample to be only $\approx$2\%.

\subsection{The $\zeta$ values of X-ray loud AGN}

X-ray data show that AGN-dominated sources populate the $\zeta\approx 0$ peak. Of the 20,039 24 ${\rm \mu m}$ sources considered in this work, most of which are likely at $z>0.6$, 1095 ($\approx$5\%) have an X-ray counterpart in the {\it Chandra} XBo\"otes survey; their $\zeta$ distribution is overplotted in Figure~\ref{fig:f24_8}. These sources occupy a narrow distribution around $\zeta\approx 0$, i.e., a flat spectrum in $\nu$f$_\nu$ consistent with the typical slope of unobscured AGN (e.g., \citealt{elv94}). Because the XBo\"otes survey has a relatively shallow flux density limit (f$_{0.5-7~{\rm keV}}\approx 7.8 \times 10^{-15} {~\rm ergs~cm^{-2}~s^{-1}}$; \citealt{ken05}), all 24 ${\rm \mu m}$ sources at $z>0.6$ with an X-ray counterpart must contain a powerful (${\rm L_x > 10^{43} ergs~s^{-1}}$) AGN. This does not mean that the mid-IR emission from 24 ${\rm \mu m}$ sources without X-ray counterparts cannot be dominated by an AGN. The X-ray emission from the AGN could be significantly attenuated in LIRGs, even in the hard X-ray band. In addition, the X-ray emission from AGN has been shown to be highly variable with time (e.g., \citealt{pao04}). Inclusion of the X-ray data serves purely to demonstrate that the $\zeta$ distribution of 24 ${\rm \mu m}$ sources hosting a powerful AGN peaks at $\zeta\approx$0. 

\subsection{The $\zeta$ values of sources with optical spectroscopy}

\begin{figure}
\plotone{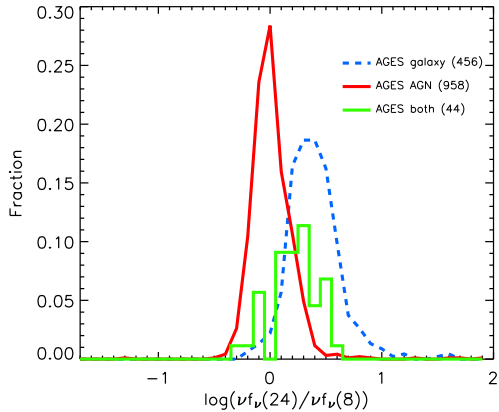}
\figcaption{\label{fig:f24_8_ages} Histograms showing $\zeta$ for all 24 ${\rm \mu m}$ sources with spectroscopic redshifts $z>0.6$ from AGES which show signs of AGN activity in their optical spectra (solid red line), which show no signs of AGN activity in their optical spectra (dashed blue line), and which show both evidence of AGN activity and stellar absorption features in their optical spectra (solid green histogram). The number of sources contributing to each histogram is listed in the figure.}
\end{figure}

Of the 20,039 24 ${\rm \mu m}$ sources at $z>0.6$, we have optical spectra for 1702 optically bright sources from AGES and 174 optically fainter sources from KECK/DEIMOS. For all optical spectra for which the signal-to-noise ratio was sufficient to perform the analysis, we classified the AGES sources by eye into those with signs of AGN activity (i.e., clear broad line emission lines or high ionization lines), only galaxy signatures, or both. Figure~\ref{fig:f24_8_ages} shows that the sources with AGN signatures fall in a narrow distribution around $\zeta\approx$0 where we expect AGN-dominated sources. The $\zeta$ distribution of sources with no AGN signatures in their optical spectra peaks at $\zeta\approx$0.5. There is a smaller tail at higher $\zeta$ values than for the general population. This is presumably because the AGES sources are all optically bright and hence do not include the most dusty, obscured and/or high redshift starburst galaxies which tend to have the highest $\zeta$ values. The sources with both AGN and galaxy signatures have a large range of $\zeta$ values, suggesting that they are powered by a combination of AGN and starburst activity. 

\begin{figure}
\plotone{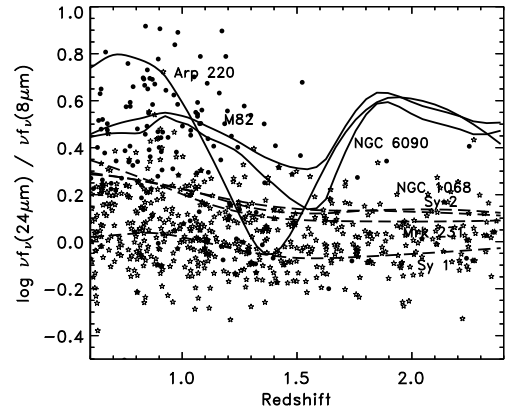}
\figcaption{\label{fig:f24f8_z} $\zeta\equiv{\rm log_{10}}[\nu f_\nu(24{\rm \mu m})/\nu f_\nu(8{\rm \mu m})]$ as a function of redshift for all $z>0.6$ sources for which we have spectroscopic redshifts from KECK/DEIMOS (174; filled circles) and for all X-ray sources from the optically bright AGES survey (616; empty stars). Overplotted are LIRG templates which are known to be starburst-dominated (solid lines: Arp 220, NGC 6090, M82; \citealt{sil98}), and AGN-dominated (dashed lines: NGC 1068; \citealt{lef01}, Mrk 231; \citealt{ega04}, averaged Sy1/2 SEDs; \citealt{cha00}).}
\end{figure}

Figure~\ref{fig:f24f8_z} shows $\zeta$ as a function of redshift for all 174 $z>0.6$ sources for which we have spectroscopic redshifts from KECK/DEIMOS and for all 616 optically bright AGES sources which have X-ray counterparts from the XBo\"otes survey. Overplotted are LIRG and ULIRG templates whose infrared emission is known to be dominated by either a starburst (Arp 220, M82, NGC 6090) or an AGN (NGC 1068, Mrk 231). The templates show that we expect starburst-dominated LIRGs to have significantly higher values of $\zeta$ (with $\zeta\approx 0.5)$ than AGN-dominated LIRGs at all redshifts except $z\sim1.1-1.6$. The pronounced dip in $\zeta$ of starburst-dominated LIRG templates is due to a strong 9.7 $\rm \mu m$ silicate absorption feature entering the observed frame 24 $\rm \mu m$ band. The possible effects of this are discussed in Section~\ref{sec:dis}. The X-ray sources span a large redshift range and have $\zeta$ values that are, in almost all cases, consistent with the AGN-dominated templates. The sources with high $\zeta$ tend to have lower redshifts, perhaps suggesting that the majority of $z>1.2$ starburst-dominated LIRGs are too faint to be included in our 24 $\rm \mu m$ sample (see, e.g., \citealt{gru05} figure~5.), have lower values of $\zeta$ due to silicate absorption, or have only weak spectral features in the observed-frame optical band. 

\subsection{The fraction of AGN-dominated LIRGs as a function of f$_{24}$.}

Assuming that the mid-IR emission from all $z>0.6$ sources in the $\zeta\approx$0 and $\zeta\approx$0.5 peaks are respectively AGN- and starburst-dominated, we can estimate the AGN-dominated fraction as a function of f$_{24}$. The number of sources in each peak were estimated by fitting the positions, widths, and amplitudes of two Gaussian profiles (chosen for simplicity and because they fit the data reasonably well). The central position of the AGN peak was determined first by fitting a Gaussian profile to the $\zeta$ histogram of the X-ray sources. The central position of the starburst peak was then determined by fitting 2 Gaussian profiles to the $\zeta$ histogram of the 0.3 mJy$~\le~$f$_{24}<~$0.5 mJy sources while fixing the position and width of the AGN peak to their estimated values. The widths and amplitudes of the two peaks were then simultaneously fit as a function of f$_{24}$, with their positions fixed. Table~\ref{tab:hist_par} lists the resulting fits. The fraction of sources whose mid-IR emission is dominated by an AGN is estimated from the area under the $\zeta\approx 0$ Gaussian divided by the total area. 

\begin{deluxetable}{llllll}
\tabletypesize{\scriptsize}
\tablecolumns{4} 
\tablewidth{0pc} 
\tablecaption{\label{tab:hist_par} Gaussian fit parameters and AGN fraction for $z>0.6$ sources.} 
\tablehead{
 \colhead{f$_{24}$}    &  \multicolumn{2}{c}{$\zeta=0.03$ peak} &  \multicolumn{2}{c}{$\zeta=0.47$ peak} & AGN \\ 
\colhead{(mJy)} & \colhead{Width$^1$} & \colhead{Amp$^2$} & \colhead{Width$^1$} & \colhead{Amp$^2$} & fraction
}
\startdata 
X-ray  & 0.17 & 117.8$\pm$4.3& -    & -             & -            \\
0.3-0.5& 0.24 & 98.7$\pm$4.5 & 0.31 & 705.0$\pm$5.4 & 0.09$\pm$0.01\\
0.5-0.7& 0.18 & 89.6$\pm$6.1 & 0.25 & 251.5$\pm$6.0 & 0.20$\pm$0.03\\
0.7-1.2& 0.16 & 61.3$\pm$4.0 & 0.20 & 73.3$\pm$3.5  & 0.38$\pm$0.06\\
1.2-1.9& 0.15 & 31.2$\pm$2.9 & 0.20 & 17.0$\pm$1.8  & 0.57$\pm$0.14\\
$>$1.9 & 0.16 & 22.9$\pm$2.2 & 0.32 & 4.1$\pm$0.9   & 0.74$\pm$0.21
\enddata
\tablenotetext{1}{The 1$\sigma$ width of the fitted Gaussian. The uncertainties are $<$0.01}
\tablenotetext{2}{Amplitude of the fitted Gaussian}

\end{deluxetable}

\begin{figure*}
\plottwo{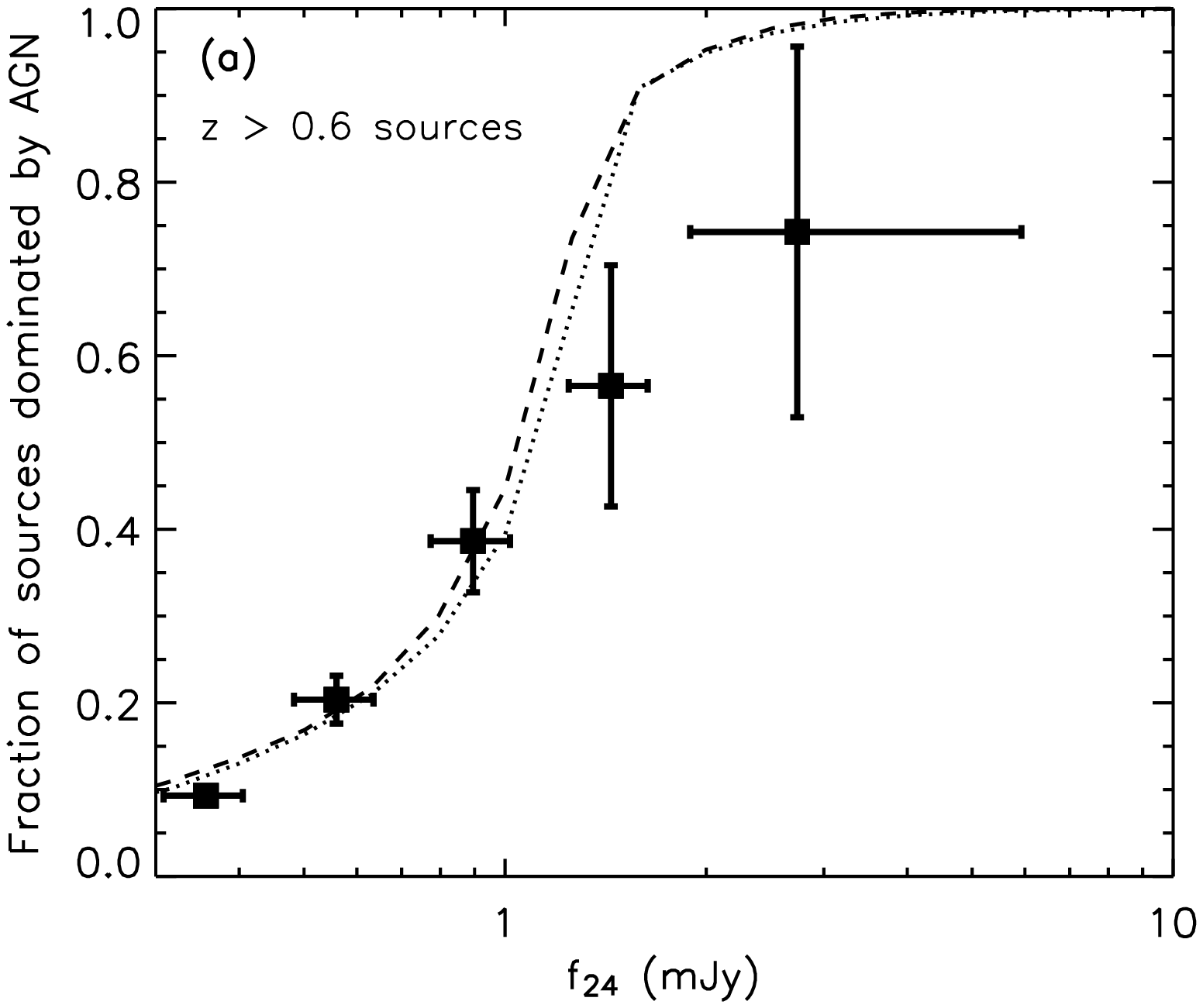}{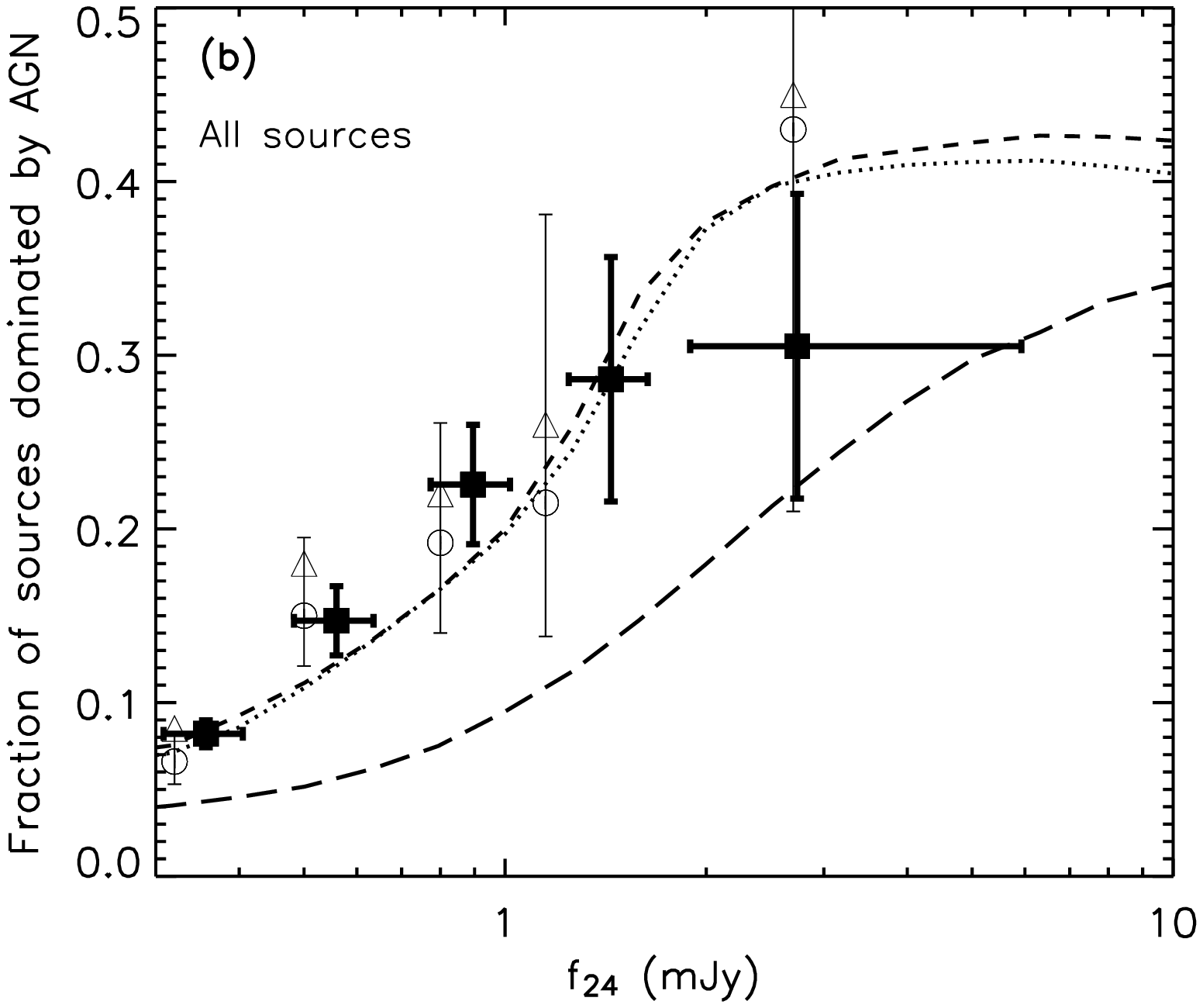}
\figcaption{\label{fig:agnfrac} The fraction of (a) all $z>0.6$ sources and (b) all 24 ${\rm \mu m}$ sources (assuming all $z<0.6$ sources are starburst-dominated) whose mid-IR emission is dominated by AGN as a function of f$_{24}$ (filled squares). The vertical error bars are calculated from the uncertainties in the Gaussian fit parameters and do not include the uncertainties introduced by the possible inclusion of a small number of optically faint $z \le 0.6$ sources or the possibility that some sources may contain a heavily embedded AGN which may be obscured even at 8 ${\rm \mu m}$. The horizontal error bars show the uncertainty on the mean f$_{24}$ in each bin. Overplotted are the expected fraction of AGN-dominated sources for the bright-end and burst models of \citet{pea05} (dotted and dashed lines respectively) and in (b), the model of \citet{gru05} (long dashed line). Also overplotted in (b) are the results of \citet{tre05} using a hard X-ray AGN classification in the GOODS fields. The points are uncorrected (empty circles) and corrected (empty triangles) for AGN not detected in the X-ray.}
\end{figure*}

Figure~\ref{fig:agnfrac}a shows that at the brightest f$_{24}$, 74$\pm$21\% of all $z>0.6$ sources have their mid-IR emission dominated by an AGN. At the faintest f$_{24}$ probed by our survey ($\approx$0.3 mJy), this fraction decreases to $\approx$9\%. Thus, in the deepest 24 ${\rm \mu m}$ surveys, starbursts will dominate the mid-IR emission in the vast majority of sources. We note that this is a measurement constrained by the assumption that the distributions of the AGN and non-AGN populations as a function of $\zeta$ can be approximated by Gaussian distributions. Figure~\ref{fig:agnfrac}b shows the fraction of all 24 ${\rm \mu m}$ sources whose mid-IR emission is dominated by an AGN as a function of their 24 ${\rm \mu m}$ flux. We assume that all $z<0.6$ sources are starburst-dominated unless they have X-ray counterparts in the XBo\"otes survey, so this calculation should be treated as a lower limit. \citet{pea05} and \citet{gru05} present phenomenological models for the evolution of different populations to determine their contributions to the 24 $\rm \mu m$ source counts. Although both models reproduce the observed trend of a higher fraction of AGN-dominated LIRGs at high f$_{24}$, the models of \citet{pea05} are in better agreement with our observations. Our results agree well with the results of \citet{tre05} who identify AGN-dominated sources by their hard X-ray emission for all 24 ${\rm \mu m}$ sources in the GOODS field (see Figure~\ref{fig:agnfrac}b).

\subsection{The AGN contribution to the 24 ${\rm \mu m}$ background.}

To estimate the contribution of AGN to the 24 ${\rm \mu m}$ background, the $z\le0.6$ sources were again included, and assumed to be starburst-dominated unless they were X-ray sources.
These results were combined with the differential number counts from \citet{pap04} to estimate the contribution of AGN to the 24 ${\rm \mu m}$ background. Down to the f$_{24}$=0.3 mJy limit of our sample, where we resolve $\sim$ 30\% of the 24 ${\rm \mu m}$ background, AGN contribute $\approx$10\% of the total mid-IR flux. This is consistent with the results of \citealt{fad02} and \citealt{fra05}. To estimate the contribution of AGN to the total 24 ${\rm \mu m}$ background, our estimates of the fraction of sources dominated by AGN were extrapolated to f$_{24}$=0.033 mJy (the flux density limit of the data used by \citealt{pap04}). \citet{pap04} estimate that they resolve $\approx$70\% of the background. We assume the remaining 30\% of the background is comprised of f$_{24}<$0.033 mJy starbursts. Assuming that the fraction of AGN-dominated sources tends to 0 and 0.1 at f$_{24}$=0.033 mJy, we find that AGN contribute 3 and 7\% of the background respectively. The 24 ${\rm \mu m}$ background is dominated by sources with low 24 ${\rm \mu m}$ flux (peaking at $\approx$0.2 mJy; \citealt{pap04}). Although AGN do not dominate the mid-IR emission in these sources, they could still contribute to a significant fraction of the total emission, and therefore make up a larger fraction of the total 24 ${\rm \mu m}$ background than the above calculation suggests. In addition, the AGN light could remain obscured in the observed 8 ${\rm \mu m}$ band but still dominate the emission at longer wavelengths. Our result should therefore be considered a lower limit.

\section{Discussion}
\label{sec:dis}

There is a clear bimodal distribution in the $\zeta$ values of $z>0.6$ 24 ${\rm \mu m}$ sources at $\zeta\approx 0$ and $\zeta\approx 0.5$. In addition, the sources with brighter f$_{24}$ are increasingly likely to populate the $\zeta\approx 0$ peak. We suggest that this bimodal behavior may be due to the presence or absence of a powerful AGN which heats the surrounding dust to high temperatures ($\approx$ 1500 K at the dust-sublimation radius; e.g., \citealt{san89}), resulting in a shallower power-law continuum slope for the AGN-dominated sources. Our confidence in this interpretation is boosted by the remarkably narrow $\zeta$ distribution of the X-ray sources at $\zeta\approx 0$. 

Making the crude assumption that, on average, the  $z>0.6$ sources with higher f$_{24}$ correspond to sources with higher IR luminosities, our results imply that the fraction of AGN-dominated sources increases significantly with IR luminosity. This is in broad agreement with a number of earlier studies at low redshift.  \citet{lut98} used {\it ISO} to study a sample of brighter, nearby ULIRGs. They measured the ratio of the 7.7 ${\rm \mu m}$ PAH emission feature to the local continuum to determine that while only 15\% of LIRGs at luminosities below 2$\times 10^{12} L_{\odot}$ are AGN powered, this number increases to $\sim$50\% at higher luminosities. 

Although the $\zeta$ diagnostic appears to work well in distinguishing AGN and starburst dominated sources, we note several caveats with its use. Many broad emission and absorption line features are known to be present in the mid-IR spectrum of LIRGs (e.g., \citealt{hou05}; \citealt{yan05}) and this may affect $\zeta$ as a function of redshift. 
We have already discussed how the 7.7 $\rm \mu m$ PAH emission feature depresses $\zeta$ when it passes through the 8 $\rm \mu m$ band in $z<0.6$ galaxies. At higher $z$, the most notable broad feature which may affect $\zeta$ is the silicate absorption feature at 9.7${\rm \mu m}$. If present, this may strongly attenuate the observed 24 ${\rm \mu m}$ emission at z$\approx$1.1-1.7, resulting in a lower $\zeta$ and hence mimicking a shallower continuum slope (see Figure~\ref{fig:f24f8_z}). Dusty starburst-dominated LIRGs at z$\approx$1.1-1.7 may therefore be mis-classified as AGN (e.g., \citealt{dei05}). 
This should not significantly affect our AGN-fraction determination since, if the absorption feature is strong enough, only a small number of affected sources will be bright enough to be selected in our shallow 24 ${\rm \mu m}$ survey. \citet{pea05} predict $\approx$25 non-AGN sources per square degree with $f_{24}>$0.25 mJy at $z\approx$1.1-1.7. We therefore expect a maximum of only $\approx$225 starburst-dominated sources ($\approx$1\% of our $z>0.6$ sample) to be mis-classified as AGN-dominated sources due to the silicate absorption feature.

It is also possible that an AGN could be heavily embedded in large amounts of cooler dust and could remain undetected in the 8 ${\rm \mu m}$ band even though it dominates the bolometric infrared emission. \citet{alo03} and \citet{cla00} show that nearby Seyfert 2 galaxies tend to have steeper mid-IR spectra than nearby Seyfert 1 galaxies, suggesting that high absorption blocks a large fraction of the mid-IR emission from the inner torus, at least in low luminosity AGN in the local Universe. In the X-ray detected AGN, we notice a weak trend of lower $\zeta$ values in the sources with fainter optical counterparts, suggesting that obscuration may influence our results. Ideally, one would estimate the continuum slope at longer wavelengths to avoid this problem. Nevertheless, although these potential effects may be important, they do not significantly dilute the bimodal nature of the $\zeta$ distribution and therefore cannot be the dominant effect.

Obtaining complete samples of optically fainter mid-IR sources with spectroscopic redshifts will be important in testing the $\zeta$ diagnostic and in drawing conclusions about the physical nature of high redshift LIRGs. In particular, this will allow us to confirm whether the fraction of AGN-dominated sources increases with IR luminosity at high redshift, and to determine the effects of measuring the dust continuum slope at different rest-frame wavelengths (can powerful AGN really be so embedded that their near-IR emission is absorbed?). Redshift information will also be crucial in determining the contribution of LIRGs to the star-formation and SMBH accretion histories in massive galaxies and in investigating whether the AGN- and starburst-dominated LIRGS and ULIRGs form an evolutionary sequence. \\

We thank our colleagues on the NDWFS, MIPS, IRS, IRAC, XBo\"otes, and AGES teams. This research is supported by the National Optical Astronomy Observatory which is operated by the Association of Universities for Research in Astronomy, Inc. (AURA) under a cooperative agreement with the NSF. We thank Chris Pearson for providing data for his models. This work is based on observations made with the {\it Spitzer Space Telescope}, which is operated by the Jet Propulsion Laboratory, California Institute of Technology under a contract with NASA. Support for this work was provided by NASA through an award issued by JPL/Caltech. Partial support for this work was provided by NASA through the Spitzer Fellowship Program, through a contract issued by the JPL, Caltech under a contract with NASA. The {\it Spitzer/MIPS} survey of the Bo\"otes region was obtained using GTO time provided by the {\it Spitzer} Infrared Spectrograph Team (James Houck, P.I.) and by M. Rieke. Spectroscopic observations of the AGES project were obtained at the MMT Observatory, a joint facility of the Smithsonian Institution and the University of Arizona. Partial support for this work was provided by the National Aeronautics and Space Administration through Chandra Award Number GO3-4176. We thank the referee for her useful comments.

\end{document}